# Two-Phase Switched Reluctance Motors: Optimal Magnet Placement and Drive System for Torque Density

Gholamreza Davarpanah, *Member, IEEE,* Sajjad Mohammadi, *Member, IEEE,* James L. Kirtley*, Life Fellow, IEEE*

*Abstract*—This paper focuses on designing new motors with high torque density, which is crucial for applications ranging from electric vehicles to robotics. We propose a double-teeth C-core switched reluctance motor with hybrid excitation, integrating permanent magnets and a novel drive technique to enhance motor torque density. We explore three magnet placement configurations to maximize torque. A common challenge with most self-starting methods used in two-phase SRMs is the generation of negative torque, which reduces the motor's torque density. Our adopted self-starting method minimizes negative torque, and we introduce a new drive strategy to control the switching on and off, effectively eliminating negative torque. Additionally, magnetic equivalent circuits are developed for the analytical design and theoretical analysis of all configurations. The SRMs under study are prototyped and tested, and their performances are evaluated in terms of torque-angle characteristics, current, and voltage. Both experimental and simulation results validate the effectiveness of the PM-assisted SRMs in enhancing torque density and efficiency.

*Index Terms*— C-core, Double-Teeth, Hybrid excitation, Magnetic equivalent circuit, permanent magnet, Switched reluctance motor.

## I. INTRODUCTION

Switched reluctance motors (SRMs) have garnered considerable attention in various applications such as electric vehicles [1], home appliances [2], and general industrial tools [3]-[5]. Their notable characteristics include a simple and robust structure, high reliability, the absence of windings and permanent magnets (PM) on the rotor, low production costs, and a wide operational speed range [6]-[10]. Introducing new structures for switched reluctance motors that focus on higher efficiency and thereby improving overall effectiveness is imperative [11]-[15]. Additionally, in the design process of SRMs, the radial magnetic forces should be carefully considered to optimize the device's performance [3].

The performance of SRMs is often compromised by high magnetic saturation in both the stator and rotor steels, as well as the nonlinear flux characteristics. Various approaches have been suggested to mitigate these drawbacks. Among them is the adoption of multiple teeth per stator pole, aimed at improving torque characteristics [16]-[20]. Another effective strategy to enhance output torque involves increasing the number of rotor poles to expand the slot area for additional turns in each phase [21], although this introduces several challenges. However, an increased number of teeth per stator pole does not consistently enhance torque. Studies shown in [18]-[20] indicate that a double-teeth structure is the most effective topology.

Despite extensive research on enhancing the torque capability of PM-Assisted motors [22]-[25], cogging torque remains an inherent challenge, and the substantial use of PMs elevates costs. Flux-switching permanent magnet motors (FSPMs), which embed PMs in the stator poles, offer a simple and robust rotor structure but suffer from high magnetic saturation in the stator. This design also reduces winding space, consequently diminishing the motor's torque capability [26], [27]. Embedding PMs into switched reluctance motors (SRMs) to enhance torque density while minimizing cogging torque has been explored in [28], [29]. An innovative SRM design is investigated, featuring supplementary windings and strategically embedded PMs within the stator yoke [28], demonstrating significant improvements in torque and efficiency. Additionally, a unique modular SRM with an unconventional A-type stator layout is introduced in [29], incorporating PMs within the stator back iron, which effectively reduces torque ripple.

Furthermore, in [30], a 6/5-pole modular hybrid reluctance motor (HRM) is meticulously designed, strategically placing PMs between poles within each module, showcasing its superiority over conventional 6/4 and 12/8 SRMs. In [31], a multi-tooth hybrid excited switched reluctance motor (MT-HESRM) integrates PMs between the end teeth of adjacent modules to enhance torque and efficiency. However, this design overlooks the impact of radial magnetic forces, which can cause deformation of the stator core, breakdown of bearings, and excessive noise during extended operation. The enhanced-torque SRM investigated in [32], featuring two sets of PMs embedded in the stator yoke and end teeth of adjacent modules, similarly neglects radial magnetic forces. Efforts to mitigate these forces include incorporating structures where each stator pole comprises three [33] and four teeth [34], though this leads to narrower rotor poles, causing saturation, higher core losses, and necessitating faster switching frequencies at constant rotor speeds, ultimately increasing losses and reducing overall efficiency. Ding [35] introduces a modular stator HRM with twelve stator poles and eight rotor poles, utilizing six PMs, yielding higher torque and power compared to traditional SRMs. An E-core hybrid reluctance motor in [36] features a novel configuration with four PMs in the common pole of the motor's stator, which consists of eight poles and four common poles, while the rotor has sixteen salient poles. This HRM demonstrates improved efficiency and reduced cogging torques. Despite these advantages, the integration of PMs within the stator teeth leads to the development of cogging torque, a significant drawback for SRM operation, as it lacks designated paths for the PM flux in the absence of rotor movement.

The main contribution of this paper is to investigate double-teeth hybrid switched reluctance motors (DT-HSRM) incorporating permanent magnets and a novel drive technique designed to counteract the negative torque often generated by

---

Gholamreza Davarpanah is with the Department of Electrical Engineering, Amirkabir University of Technology, Tehran 158754413, Iran (e-mail: ghr.davarpanah@aut.ac.ir).

Sajjad Mohammadi and James L. Kirtley are with the Massachusetts Institute of Technology, Cambridge, MA 02139, USA (e-mail: sajjadm@mit.edu; kirtley@mit.edu).

conventional starting methods, thereby significantly enhancing the motor's torque density. The PMs are strategically embedded between the teeth of the C-core poles to optimize magnetic interactions. A common challenge with two-phase SRMs is the unpredictability of the rotation direction owing to the absence of inherent starting torque. Traditional methods often involve creating an asymmetric air-gap by modifying the rotor poles [37]-[38]. This study leverages a refined self-starting technique outlined in [39] and proposes a control strategy that ensures the production of only positive torque. Section II elucidates the operational principles and magnetic equivalent circuits (MEC) of the investigated motor configurations. Section III discusses the simulation, analysis, and comparative evaluation of the DT-HSRMs. Section IV presents the experimental results obtained from prototype testing. The findings and conclusions are summarized in Section V.

## II. MOTOR STRUCTURES AND PREDICTIONS OF THE FLUX PATHS

In this paper, we explore a topology for SRMs that features two teeth per C-core pole and incorporates embedded permanent magnets (PMs). The motor operates at a nominal speed of 600 rpm and delivers approximately 60 watts of power. While a 60-watt motor with a speed of 600 rpm may not meet the demands of transportation electrification, this design can be scaled up for higher power applications. Additionally, it holds potential for use in small-scale electric mobility devices where lower power motors are adequate.

### A. SRM Structures

The investigated structures incorporate two C-cores in each phase, with each C-core pole split into two teeth [40], [41]. These C-cores of the same phase are positioned on the opposite sides of the rotor to cancel out radial magnetic forces, which are primary contributors to acoustic noise and stator core deformation. Overall, the structure comprises four C-cores, resulting in sixteen stator teeth. In line with its two-phase design, the rotor includes eighteen teeth. Additionally, the design includes two sets of PMs, each set containing four magnets, ensuring a continuous opposition between the magnetic flux produced by the PMs and that generated by the excited coils. The first set of PMs creates flux in a clockwise direction, while the second set induces a counterclockwise flux. Fig. 1 showcases four motor configurations designed to enhance torque production: a double-teeth two-phase switched reluctance motor (DT-SRM) and three double-teeth two-phase hybrid reluctance motors (DT-HRMs).

In the concentrated-wound stator configuration, coils from each phase are connected in series, encompassing four coils per phase. Fig. 1(a) displays the DT-SRM configuration featuring four C-cores without PMs [23]. Fig. 1(b) introduces the first DT-HRM, identical to the DT-SRM but incorporating four PMs (set 1) positioned between the internal teeth of the C-cores to generate a clockwise magnetic flux. Fig. 1(c) presents the second DT-HRM, which retains the DT-SRM structure and adds four PMs (set 2) [23] positioned between the side teeth of adjacent C-cores to produce a counterclockwise flux. Finally, Fig. 1(d) illustrates the last DT-HRM, mirroring the DT-SRM layout and including eight PMs (comprising both set 1 and set 2), where the flux path of the PMs systematically counters the flux from the excited coils in the stator teeth. This arrangement offers several advantages over conventional SRMs:

**1) Shorter flux path:** Using the C-core topology [40], [41], the magnetic flux closes its path inside the C-cores, leading to a shorter flux path, decreasing core losses, and increasing the mean torque, output power, and efficiency.

**2) Higher number of teeth at each C-core pole:** An approach to increase mean torque is to have multiple teeth per stator pole [16]-[19]. The co-energy can be calculated as follows [18]-[19]:

$$W_m = \int \frac{N^2 I \mu_0 m l_{arc} L}{2 l_g} di = \int \frac{N^2 I \mu_0 m \overbrace{((D/2)\theta)}^{l_{arc}} L}{2 l_g} di \quad (1)$$

where $N$, $I$, $\mu_0$, $m$, $l_{arc}$, $L$, $D$, and $l_g$ represent the number of turns per phase, phase current, the permeability of free space, number of teeth at each C-core pole, overlapping length between C-core teeth and rotor poles, stack length, stator inner diameter, and air-gap length, respectively. Also, the electromagnetic torque is obtained as:

$$T_e = \frac{\partial W_m}{\partial \theta} = m \cdot \int \frac{N^2 I \mu_0 D L}{4 l_g} di \quad (2)$$

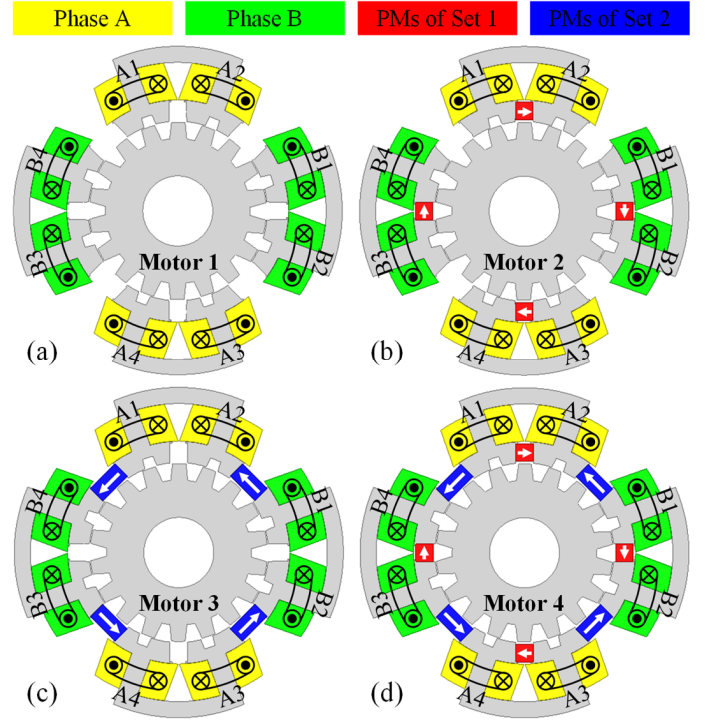

Fig. 1. The structure of the proposed motors. (a) SRM with a double-teeth per C-core pole. (b) DT-HRM with PMs positioned between the internal teeth of the C-cores. (c) DT-HRM with PMs between the side teeth of adjacent C-cores. (d) DT-HRM incorporates both sets of PMs.

Therefore, the electromagnetic torque increases with the increasing number of teeth at each C-core pole. As $m$ goes up, the rotor pole count, $N_r$, also rises. This results in narrower rotor poles, leading to increased core loss, reduced commutation angle, and a smaller winding slot area [33], [34]. Article [18] explores torque and core loss of motors with different numbers of teeth per stator pole under identical excitation currents, showing that the optimal number of teeth per stator pole is two.

**3) PM Utilization:** The electromagnetic torque in HRMs goes up compared to conventional SRMs [28]- [30].

**4) Flux density reduction in laminations:** The flux generated by embedded PMs is in the opposite direction of the flux produced by the excited windings, which results in a decrease in the flux density within the laminations [19], [35].

**5) Increasing the flux density in the air-gap:** The flux added by PMs to the air gap contributes to improving the electromagnetic torque of the motor [19], [35].

**6) High reliability:** The PMs in permanent magnet synchronous motors (PMSMs) are placed on the rotor, whereas in the investigated HRMs, the PMs are embedded inside the stator, which helps with the robustness and reliability of the motor [6]-[10].

**7) Minimum radial magnetic forces on the stator:** Radial forces are canceled out, enhancing the lifetime of the motor.

**8) Self-starting capability:** This paper utilizes the self-starting technique outlined in [39] along with a new control strategy to enhance the performance of the motor.

While the aforementioned concepts may not be new individually, the amalgamation of these features represents an innovative topology. The specifications of the studied motors, shown in Fig. 1, are detailed in Table I.

### B. Flux Paths

Fig. 2 depicts the magnetic flux paths within the studied SRMs. As illustrated in the first column of Fig. 2, without excitation current, the magnetic flux generated by the PMs travels through the stator modules without crossing the air-gap, resulting in minimal cogging torque. In contrast, as shown in the second column of Fig. 2, when a phase is excited, saturation occurs in the C-core, causing the PM flux to be directed across the air-gap into the rotor, enhancing torque production. The higher the current, the larger the saturation, and consequently, the greater the torque generated as an increased amount of PM flux crosses the air-gap.

### C. Magnetic Equivalent Circuits

As shown in Fig. 3, magnetic equivalent circuits (MEC) are developed to analytically analyze the operating principles of the SRMs. This analytical assessment ignores leakage flux, mutual flux, and saturation effect for a more straightforward study [42]-[44]. Assuming phase A is excited in all investigated motors, the reluctances of the stator yoke, stator pole, rotor yoke, air-gap, PMs of set 1, and PMs of set 2 are represented in Fig. 3 as $R_{sy}$, $R_{sp}$, $R_{ry}$, $R_g$, $R_{PM1}$, and $R_{PM2}$, respectively. The magneto motive force (MMF) of excited coils, PMs of set 1, and PMs of set 2 are denoted as $F_e$, $F_{PM1}$, and $F_{PM2}$, respectively. $\varphi_{sy}$, $\varphi_{sp}$, $\varphi_{ry}$, $\varphi_g$, $\varphi_{PM1}$, and $\varphi_{PM2}$ denote the magnetic flux of the stator yoke, stator pole, rotor yoke, air-gap, PMs of set 1, and PMs of set 2, respectively. It's important to note that the equations ($\varphi_{sp} = \varphi_{sy}$) and ($\varphi_{ry} = \varphi_g$) hold for all four topologies. The process of calculating magnetic fluxes is elaborated for Motor 4 with eight PMs (DT-HRM with both sets of PMs), and a similar approach can be applied to the other three motors. As shown in Fig. 3(d), since the reluctances of PMs of set 1 and PMs of set 2 are significantly larger than the reluctances of the iron part of the motors, the following equations are always valid.

$$\begin{cases} R_{PM1} \gg (2R_g + R_{sy}) \\ R_{PM2} \gg (2R_{sp} + R_{sy}) \end{cases} \quad (3)$$

Considering the simple approximations in equations (3), two points, A and A', are short circuits. Hence, a simpler equivalent circuit of the investigated designs can be derived, as depicted in Figs. 3(e) and 3(f). The magnetic equivalent circuits shown in Figs. 3(e) and 3(f) represent the simplified MECs for Motors 3 and 4, respectively. According to Fig. 3(f), the following fundamental equations can be derived using Kirchhoff's laws:

$$\varphi_g^A - \varphi_{sy}^A - \varphi_{PM1}^A - \varphi_{PM2}^A = 0 \quad (4)$$

$$2R_{PM2}\varphi_{PM2}^A + (2R_g + R_{ry})\varphi_g^A = 2F_{PM2} \quad (5)$$

$$(2R_{sp} + R_{sy})\varphi_{sy}^A - R_{PM1}\varphi_{PM1}^A = 2F_e - F_{PM1} \quad (6)$$

$$(2R_g + R_{ry})\varphi_g^A + R_{PM1}\varphi_{PM1}^A = F_{PM1} \quad (7)$$

TABLE I
SPECIFICATIONS OF MOTORS.

| | Motor 1 | Motor 2 | Motor 3 | Motor 4 |
|---|---|---|---|---|
| Stator outer diameter, $D_o$ (mm) | 94 | 94 | 94 | 94 |
| Stator yoke thickness, $b_{sy}$ (mm) | 4.6 | 4.6 | 4.6 | 4.6 |
| Stator pole height, $h_s$ (mm) | 10.8 | 10.8 | 10.8 | 10.8 |
| Tooth height thickness, $b_{ty}$ (mm) | 3.2 | 3.2 | 3.2 | 3.2 |
| Tooth length, $h_t$ (mm) | 2.8 | 2.8 | 2.8 | 2.8 |
| Air-gap length, $l_g$ (mm) | 0.3 | 0.3 | 0.3 | 0.3 |
| Rotor pole length, $h_r$ (mm) | 4.64 | 4.64 | 4.64 | 4.64 |
| Stator pole arc, $\lambda$ (deg) | 10 | 10 | 10 | 10 |
| Stator tooth arc, $\beta_t$ (deg) | 8.4 | 8.4 | 8.4 | 8.4 |
| Stator tooth arc, $\beta_t + \alpha$ (deg) | 11.9 | 11.9 | 11.9 | 11.9 |
| Extension of the teeth, $\alpha$ (deg) | 3.56 | 3.56 | 3.56 | 3.56 |
| Rotor pole arc, $\beta_r$ (deg) | 8.8 | 8.8 | 8.8 | 8.8 |
| Stack length, $L$ (mm) | 20 | 20 | 20 | 20 |
| PM width, $W_{PM}$ (mm) | - | 5 | 5 | 5 |
| PM length, $l_{PM1}$ & $l_{PM2}$ (mm) | - | 5 | 10 | 5 & 10 |
| Number of turns per pole of a C-core | 90 | 90 | 90 | 90 |
| Type of lamination material | M19-24G (knee point = 1.9T) | | | |
| Type of PMs | NdFeB-N35 | | | |

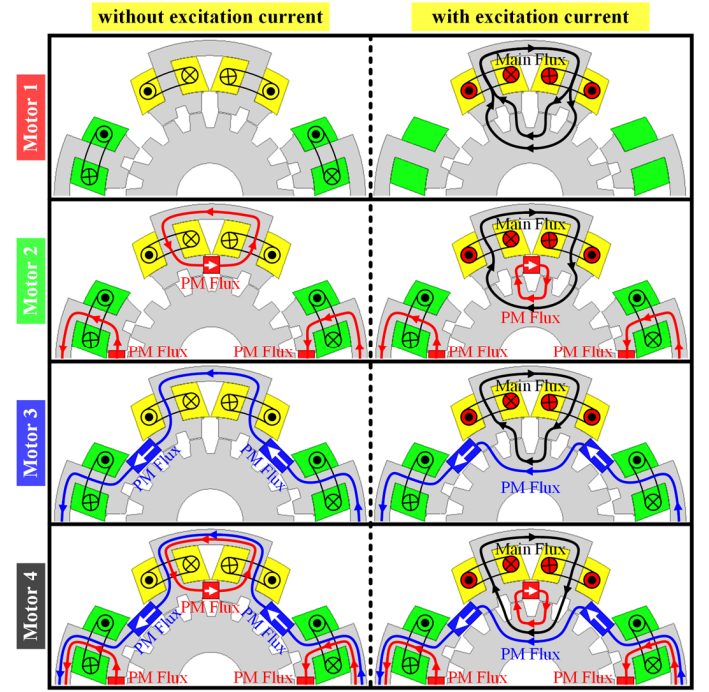

Fig. 2. Predicted magnetic flux paths of four two-phase investigated motors.

To facilitate derivations, the following equations are considered:

$$\begin{cases} R_{PM1} \gg (2R_g + R_{ry}) \\ R_{PM2} \gg (2R_g + R_{ry}) \end{cases} \quad (8)$$

$$\begin{cases} R_{PM1} \gg \left[ (2R_{sp} + R_{sy}) \| (2R_g + R_{ry}) \right] \\ R_{PM2} \gg \left[ (2R_{sp} + R_{sy}) \| (2R_g + R_{ry}) \right] \end{cases} \quad (9)$$

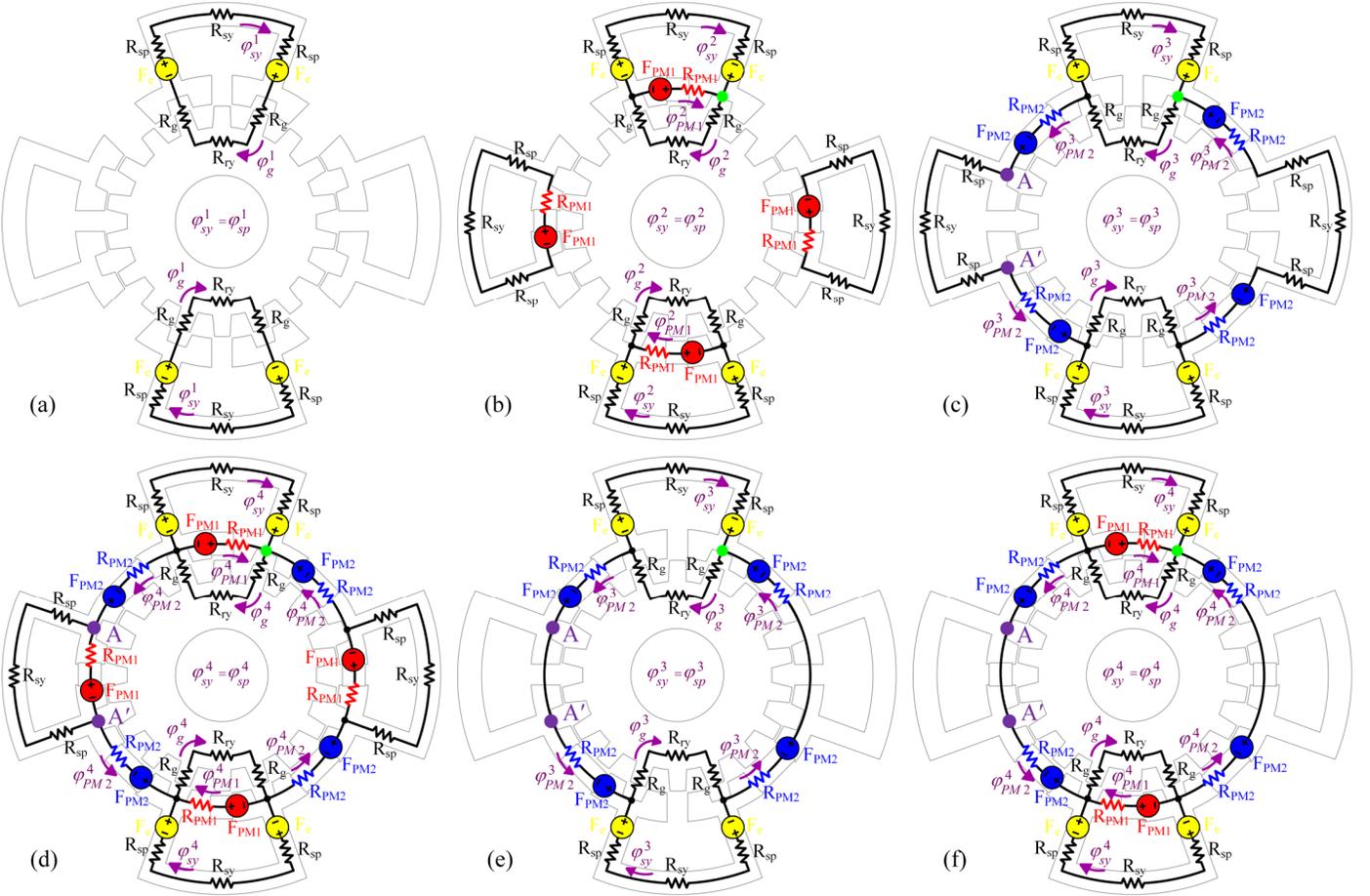

Fig. 3. Magnetic equivalent circuits for (a) Motor 1, (b) Motor 2, (c) Motor 3, and (d) Motor 4, along with simplified magnetic equivalent circuits for (e) Motor 3 and (f) Motor 4.

Magnetic fluxes within the stator yoke, stator pole, and air-gap for the Motor 4 can be derived as in shown below.

$$\varphi_{sy}^4 = \varphi_{sp}^4 = \overbrace{+\frac{1}{R^*} \times 2F_e}^{\varphi'_{sy} = \varphi'_{sp}} - \overbrace{\frac{2R_g + R_{ry}}{R_{PM1}R^*} \times F_{PM1}}^{\varphi''_{sy} = \varphi''_{sp}} - \overbrace{\frac{2R_g + R_{ry}}{R_{PM2}R^*} \times F_{PM2}}^{\varphi'''_{sy} = \varphi'''_{sp}} \quad (10)$$

$$\varphi_g^4 = \overbrace{+\frac{1}{R^*} \times 2F_e}^{\varphi'_g} + \overbrace{\frac{2R_{sp} + R_{ry}}{R_{PM1}R^*} \times F_{PM1}}^{\varphi''_g} + \overbrace{\frac{2R_{sp} + R_{ry}}{R_{PM2}R^*} \times F_{PM2}}^{\varphi'''_g} \quad (11)$$

where $R^* = 2R_g + R_{ry} + 2R_{sp} + R_{sy}$     (12)

Likewise, the fluxes for Motors 3, 2, and 1 are obtained as follows:

$$\varphi_{sy}^3 = \varphi_{sp}^3 = \overbrace{+\frac{1}{R^*} \times 2F_e}^{\varphi'_{sy} = \varphi'_{sp}} - \overbrace{\frac{2R_g + R_{ry}}{R_{PM2}R^*} \times F_{PM2}}^{\varphi'''_{sy} = \varphi'''_{sp}} \quad (13)$$

$$\varphi_g^3 = \overbrace{+\frac{1}{R^*} \times 2F_e}^{\varphi'_g} + \overbrace{\frac{2R_{sp} + R_{ry}}{R_{PM2}R^*} \times F_{PM2}}^{\varphi'''_g} \quad (14)$$

$$\varphi_{sy}^2 = \varphi_{sp}^2 = \overbrace{+\frac{1}{R^*} \times 2F_e}^{\varphi'_{sy} = \varphi'_{sp}} - \overbrace{\frac{2R_g + R_{ry}}{R_{PM1}R^*} \times F_{PM1}}^{\varphi''_{sy} = \varphi''_{sp}} \quad (15)$$

$$\varphi_g^2 = \overbrace{+\frac{1}{R^*} \times 2F_e}^{\varphi'_g} + \overbrace{\frac{2R_{sp} + R_{ry}}{R_{PM1}R^*} \times F_{PM1}}^{\varphi''_g} \quad (16)$$

$$\varphi_{sy}^1 = \varphi_{sp}^1 = \varphi_g^1 = \overbrace{+\frac{1}{R^*} \times 2F_e}^{\varphi'_{sy} = \varphi'_{sp} = \varphi'_g} \quad (17)$$

In the absence of PMs, the magnetic flux in the stator yoke, stator pole, and air-gap are represented as $\varphi'_{sy}$, $\varphi'_{sp}$, and $\varphi'_g$, respectively. Simply put, these flux values result from the magnetic field generated by the phase winding excitation. Now, with the presence of PMs from set 1, the impact of these PMs on the magnetic flux in the stator yoke, stator pole, and air-gap is denoted as $\varphi''_{sy}$, $\varphi''_{sp}$, and $\varphi''_g$, respectively. In simpler terms, these flux values result from the magnetic field generated by the PMs from set 1. Similarly, with the presence of PMs from set 2, the effect of these PMs on the magnetic flux in the stator yoke, stator pole, and air-gap is represented as $\varphi'''_{sy}$, $\varphi'''_{sp}$, and $\varphi'''_g$, respectively. These fluxes originate from the PM set 2. Now, equations (10) to (17) can be analyzed for each motor topology.

**Motor 1:** Magnetic flux is generated merely by phase windings, resulting in $\varphi'_{sy}$, $\varphi'_{sp}$ and $\varphi'_g$ (equation (17)).

**Motor 2:** The magnetic flux is generated by both the phase windings and PMs of set 1, resulting in $\varphi_{sy}$, $\varphi_{sp}$, $\varphi_g$, $\varphi''_{sy}$, $\varphi''_{sp}$, and $\varphi''_g$. As stated in equation (16), the magnetic flux within the air-gap experiences an increase as a result of the additional magnetic fluxes $\varphi''_g$. Both the magnetic flux of excited phase coils and PMs of set 1 contribute to the overall magnetic flux within the air-gap. According to equation (15), the magnetic flux of both the stator yoke and stator pole decreases due to the magnetic fluxes $\varphi''_{sy} = \varphi''_{sp}$.

**Motor 3:** Both the phase windings and PMs of set 2 produce magnetic flux, resulting in $\varphi'_{sy}$, $\varphi'_{sp}$, $\varphi'_g$, $\varphi'''_{sy}$, $\varphi'''_{sp}$, and $\varphi'''_g$. Herein, there is an analysis similar to that of Motor 2. As expressed in equation (14), the magnetic flux in the air-gap undergoes augmentation because of the supplementary magnetic fluxes $\varphi'''_g$. Additionally, per equation (13), both the magnetic flux of the stator yoke and stator pole diminishes owing to the magnetic fluxes $\varphi'''_{sy}=\varphi'''_{sp}$.

**Motor 4:** At this motor, phase windings, both PMs of set 1 and PMs of set 2 produce magnetic flux, resulting in $\varphi'_{sy}$, $\varphi'_{sp}$, $\varphi'_g$, $\varphi'''_{sy}$, $\varphi'''_{sp}$, $\varphi'''_g$, $\varphi'''_{sy}$, $\varphi'''_{sp}$, and $\varphi'''_g$. As stated in equation (11), the magnetic flux within the air-gap experiences an increase as a result of the additional magnetic fluxes $\varphi''_g$ and $\varphi'''_g$. The magnetic flux of excited phase coils, PMs of set 1, and PMs of set 2 contribute to the overall magnetic flux within the air-gap. According to equation (10), the magnetic flux of both the stator yoke and stator pole decreases due to the magnetic fluxes $\varphi'''_{sy}=\varphi'''_{sp}$ and $\varphi'''_{sy}=\varphi'''_{sp}$.

## III. Simulation Result and Comparison

### A. Self-starting capability

One of the challenges in two-phase Switched Reluctance Motors (SRMs) is the lack of starting torque, resulting in an undetermined direction of rotation. This occurs because exciting a phase winding for a full period (360 electrical degrees) produces positive torque during the first half period and negative torque during the other half period. Due to the symmetric nature of the torque-angle characteristics, there is no starting torque when one phase is aligned, and the other phase needs to be excited. To overcome this zero-torque condition, creating an asymmetry is essential. Previously, we introduced a technique in [39] to address this issue, which has been implemented in the motors studied here. In this design, one tooth of each stator pole's two teeth remains unchanged, while the width of the other tooth is slightly increased to break the symmetry in the torque-angle characteristics. Consequently, a positive torque is generated for a period exceeding 180 electrical degrees with each phase excitation. Initially, the width of the teeth is 8.34 degrees, but through sensitivity analysis within a small range, the optimal width of one tooth, including the extension, is determined to be 11.9 degrees.

### B. Proposed Drive and Control Technique

In the conventional drive and control techniques, an asymmetric air-gap is typically created by either removing a small portion from or adding a hole to one side of the rotor poles [37], [38]. However, this approach results in a considerable shift in the aligned position without any control mechanism, which leads to a negative torque and reduces the average torque. That means the positive torque period of each phase remains approximately a half period and is not increased effectively. Also, it leads to a decrease in the starting torque when the unaligned phase is excited. When employing any self-starting technique for two-phase SRMs, it is imperative to ensure that the aligned position of each phase remains unchanged; otherwise, a negative torque is produced. To put it simply, when the excited phase reaches the aligned position, the next phase, which is going to be excited, should have a positive torque (start torque). In other words, the unaligned position of the next phase should happen before the aligned position of the excited phase. The distance between these two points is defined as α

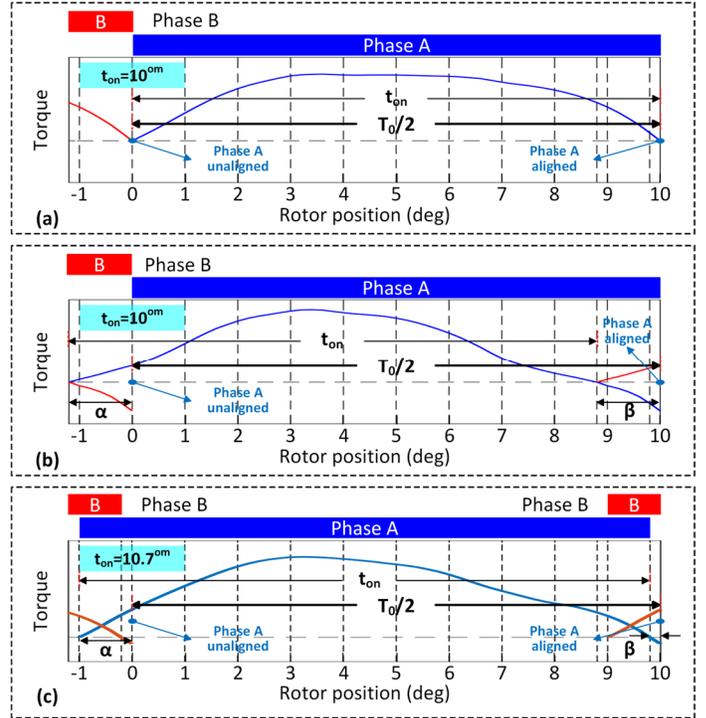

Fig. 4. Torque-angle characteristics: (a) without self-starting technique, and (b) conventional self-starting technique, and (c) proposed self-starting technique.

[39]. In the region spanning α, both phases have a positive torque, so, in the proposed control technique, the next phase can be turned on earlier by a value of α before the excited phase is turned off. This overlapping period results in an increase in the average torque. Since there is still a small shift in the aligned position using the self-starting technique, the excited phase has a small period spanning β at the end, where it produces negative torque. To eliminate this negative torque, in the proposed control technique, the excited phase is turned off sooner by a value of β. The turn on period is as follows:

Fig. 4 shows torque-angle characteristics for two-phase SRMs without self-starting torque, with conventional self-starting torque, and with proposed self-starting torque. In conventional drive and control techniques for two-phase SRMs, an asymmetric air-gap is typically created by either removing a small portion from or adding a hole to one side of the rotor poles [37], [38]. However, this approach leads to a considerable shift in the aligned position without any control mechanism, resulting in negative torque and reducing the average torque output. Consequently, the positive torque period for each phase does not effectively extend beyond half of the cycle, and this modification also diminishes the starting torque when the unaligned phase is excited. It's crucial when employing any self-starting technique for two-phase SRMs to maintain the aligned position of each phase; otherwise, negative torque is produced. Essentially, as the excited phase reaches the aligned position, the subsequent phase to be excited should generate positive torque (starting torque). This requires the unaligned position of the subsequent phase to occur before the aligned position of the currently excited phase. The distance between these two points is defined as α [39]. In the span of α, both phases produce positive torque. Therefore, in the proposed control technique, the next phase can be turn on earlier—by a value of α—before the current phase is turned off. This overlapping period enhances the average torque. Nonetheless, since the self-starting technique still creates a slight shift in the aligned position, resulting in a small interval, $\beta$, at the end of the excited phase's cycle where negative torque occurs, the proposed technique involves turning off the excited phase slightly earlier by a value of $\beta$ to avoid producing this negative torque. The turn on period is obtained as follows:

$$\theta_{on} = 180^{o} + \alpha - \beta \xrightarrow{\alpha > \beta} \theta_{on} > 180 \qquad (18)$$

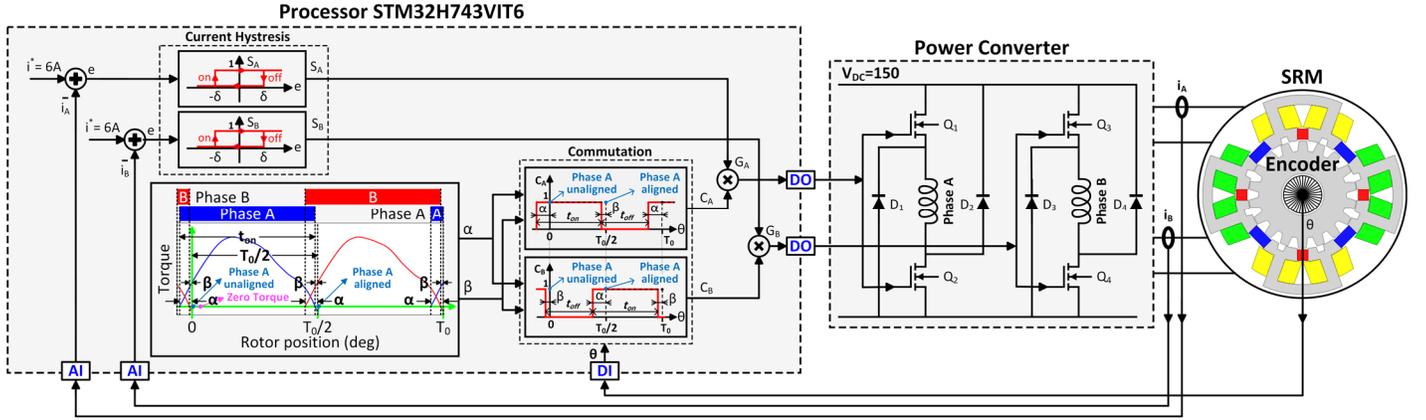

Fig. 5. Block diagram of the proposed current hysteresis control.

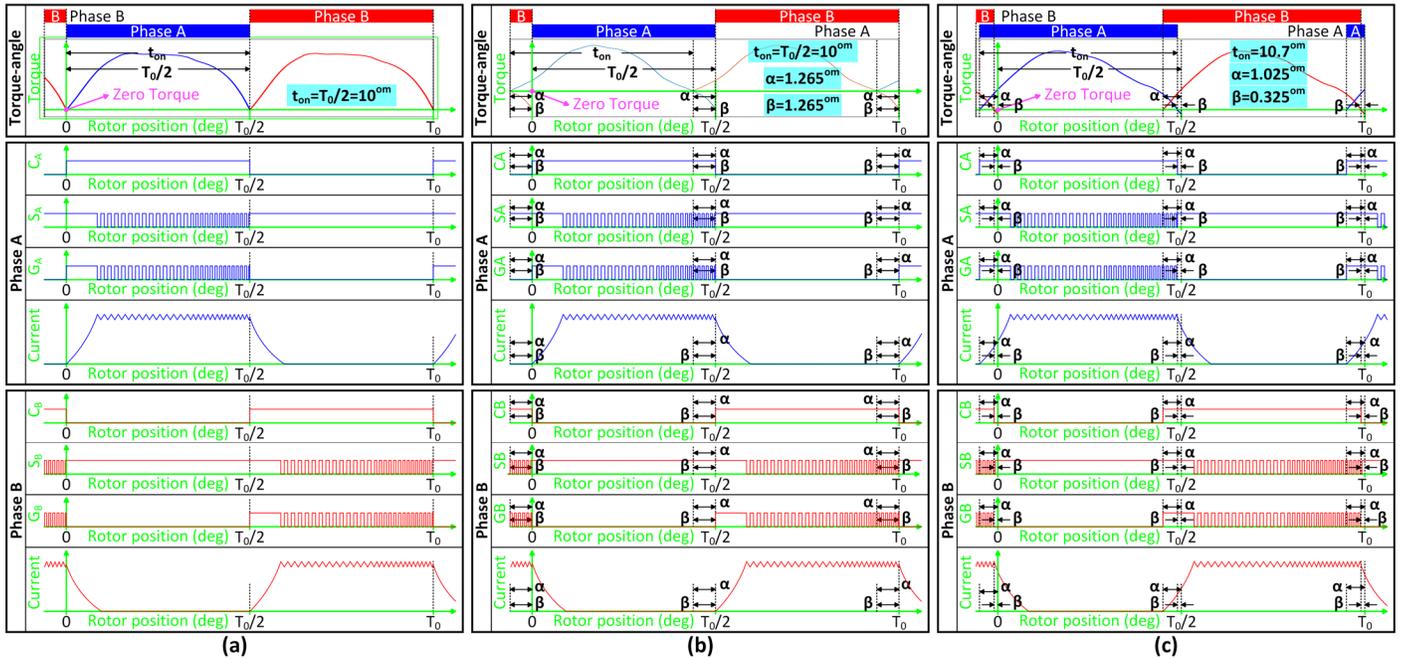

Fig. 6. Current hysteresis control: (a) SRM without self-starting torque, (b) SRM with self-starting and old control technique, and (c) SRM with self-starting torque and the proposed control technique.

The block diagram of the proposed drive system is shown in Fig. 5. It features a two-leg half-bridge inverter with free-wheeling diodes. The processor receives data from current sensors and the encoder through analog and digital inputs, respectively. After processing the data using the implemented control technique, gate pulses are generated through digital outputs. Hysteresis control compares the measured currents $i_A$ and $i_B$ with the reference current $i^*$, producing logic signals $S_A$ and $S_B$. Concurrently, commutation signals $C_A$ and $C_B$, which determine which phase to turn on, are generated using the position data and a lookup table constructed based on the values of $\alpha$ and $\beta$ obtained from the torque-angle characteristics of the motor. Finally, the gate pulses $G_A$ and $G_B$ are generated as in below:

$$G_A = S_A \cdot C_A \qquad (19)$$

$$G_B = S_B \cdot C_B \qquad (20)$$

These gate signals serve as the switch pulses for the phase that is activated. For the experimental tests, a hysteresis current reference of 6 A at a nominal speed of 600 rpm is established. Each phase remains turned on for more than 10 mechanical degrees (equivalent to 180 electrical degrees). The hysteresis band is set at $\delta$=0.2 A. Fig. 6 displays the waveforms of the control signals alongside the phase currents for three scenarios: an SRM without self-starting torque, where both $\alpha$ and $\beta$ values are zero; an SRM with conventional self-starting technique and the traditional control technique, where $\alpha$ and $\beta$ values are non-zero but identical, resulting in a phase turn-on duration close to 180 electrical degrees; and an SRM with self-starting torque and the proposed control technique, where $\alpha$ and $\beta$ values are non-zero and $\alpha$>$\beta$, leading to a phase turn-on duration exceeding 180 electrical degrees.

## C. Flux Density Distributions

Fig. 7 displays the magnetic flux density distribution within the four motors at both aligned and unaligned positions with an excitation current of 6 A. It is apparent that the presence of PMs mitigates saturation in both the stator poles and yoke. To validate the flux path predictions outlined in Section II, Fig. 8 presents the magnetic flux density vectors. It is observed that the flux from the excited windings follows a shorter path through the C-core, which contributes to reduced core losses. In contrast, the magnetic flux from the PMs enters the rotor poles via the air-gap, boosting the flux density within the air-gap and thus increasing torque. Additionally, Fig. 9 compares the air-gap flux densities of the motors at the aligned position with an excitation current of 6 A, demonstrating an enhancement in flux density due to the embedded PMs, which in turn leads to an increase in torque.

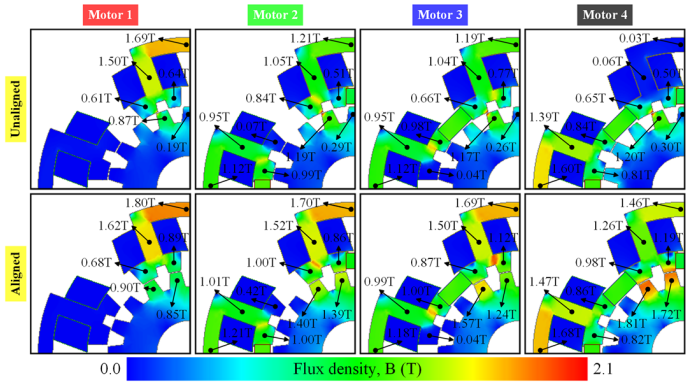

Fig. 7. Magnetic flux density distributions.

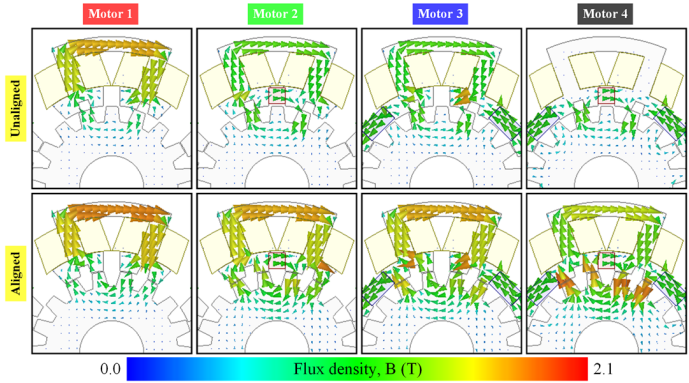

Fig. 8. Magnetic flux density vectors.

### D. Static Simulations

The torque-angle characteristics of the four motors, along with mean and peak torques for different excitation currents, are depicted in Fig. 10 and Table II, respectively. The advantages of embedding PMs in the motor structure are evident. Motor 4 delivers the highest torque. At 4 A, all motors with embedded PMs (Motors 2, 3, and 4) show nearly identical mean torque. However, as the current increases, Motor 4 surpasses Motors 2 and 3 as more PM flux enters the air-gap. Notably, Motor 2, despite having a smaller volume of PMs than Motor 3, achieves higher mean torque across all excitation currents. This is because nearly all the flux generated by PMs of set 1 in Motor 2 enters the air-gap, whereas about half of the flux generated by PMs of set 2 in Motor 3 does so. The location of set 1 proves to be more effective for PM placement. Fig. 11 shows the torque component produced by PMs, highlighting that at lower excitation currents (0-2 A), the contribution of PMs to torque generation is minimal, as most of the PM flux circulates through the C-cores. The influence of embedding PMs becomes more pronounced at higher currents (2-6 A) as more of their flux passes the air-gap due to C-core saturation by excitation currents.

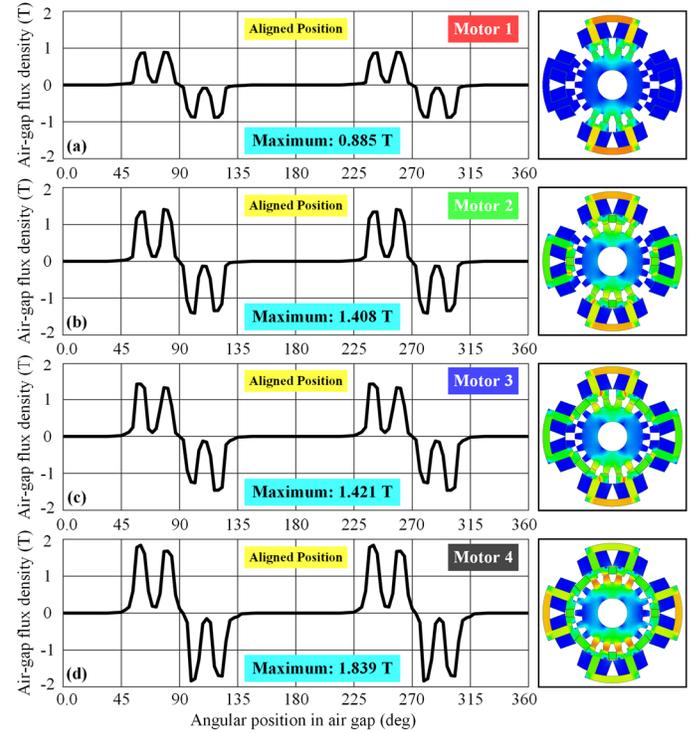

Fig. 9. Air-gap flux density comparison for four motors in aligned condition: (a) Motor 1, (b) Motor 2, (c) Motor 3, and (d) Motor 4.

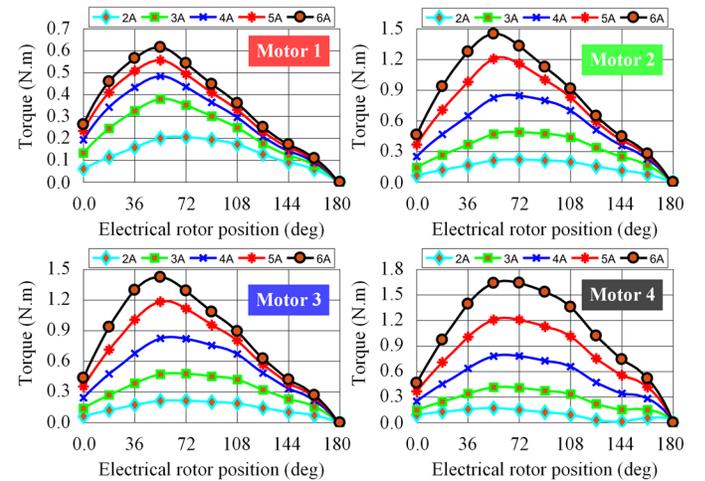

Fig. 10. Torque-angle characteristics of the four motors at the different currents.

TABLE II
Comparison Of The Mean And Peak Static Torque For Four Two-Phase Investigated Motors.

| Current (A) | Torque of Motor 1 (N.m) | | Torque of Motor 2 (N.m) | | Torque of Motor 3 (N.m) | | Torque of Motor 4 (N.m) | | Increase in mean torque of Motor 2 compared to Motor 1 (%) | Increase in mean torque of Motor 3 compared to Motor 1 (%) | Increase in mean torque of Motor 4 compared to Motor 1 (%) |
|---|---|---|---|---|---|---|---|---|---|---|---|
| | Mean Torque | Peak Torque | Mean Torque | Peak Torque | Mean Torque | Peak Torque | Mean Torque | Peak Torque | | | |
| 1 | 0.038 | 0.054 | 0.039 | 0.057 | 0.038 | 0.056 | 0.028 | 0.089 | 2.63 | 0.00 | -26.31 |
| 2 | 0.140 | 0.206 | 0.151 | 0.222 | 0.150 | 0.212 | 0.099 | 0.171 | 7.85 | 7.14 | -29.28 |
| 3 | 0.234 | 0.392 | 0.330 | 0.492 | 0.328 | 0.471 | 0.273 | 0.422 | 41.02 | 40.17 | 16.66 |
| 4 | 0.291 | 0.496 | 0.546 | 0.855 | 0.534 | 0.797 | 0.525 | 1.798 | 87.63 | 83.50 | 80.41 |
| 5 | 0.330 | 0.570 | 0.727 | 1.223 | 0.704 | 1.116 | 0.815 | 1.232 | 120.30 | 113.33 | 146.96 |
| 6 | 0.363 | 0.628 | 0.858 | 1.452 | 0.833 | 1.374 | 1.100 | 1.666 | 136.36 | 129.47 | 203.03 |

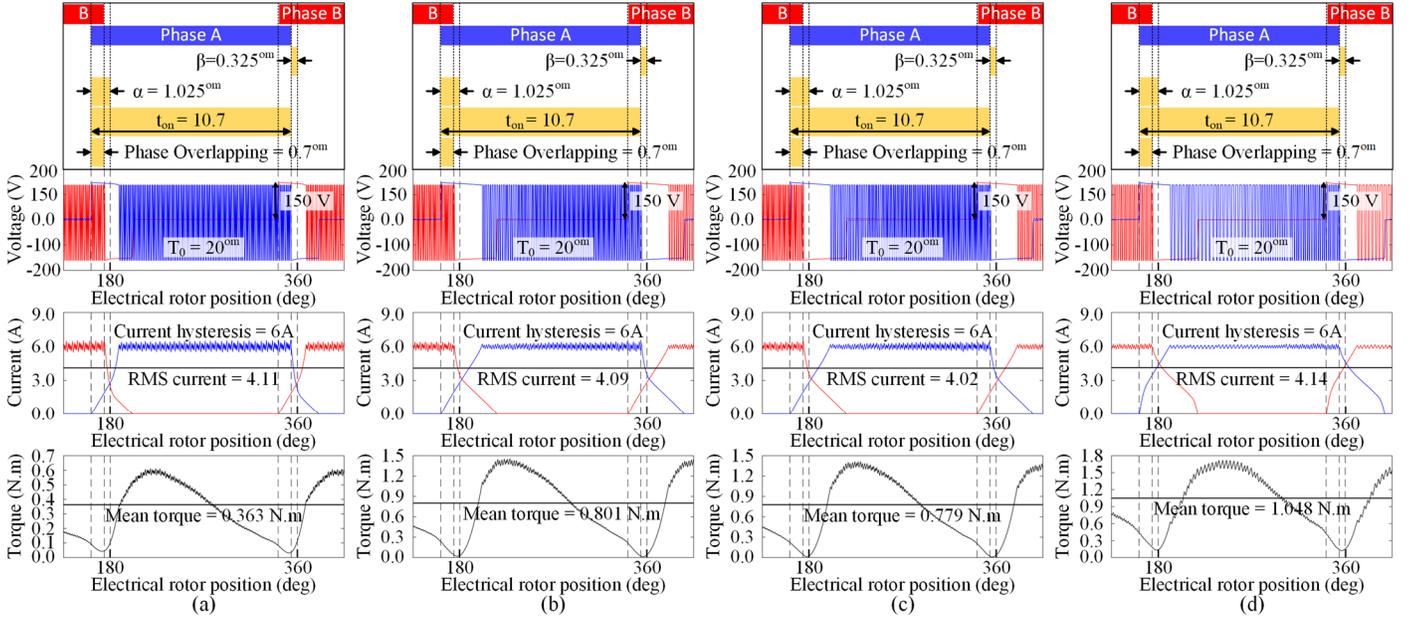

Fig. 12. Steady-state voltage, current hysteresis, and torque waveforms at the current hysteresis of 6A, 600 rpm, and CHC mode: (a) Motor 1, (b) Motor 2, (c) Motor 3, and (d) Motor 4.

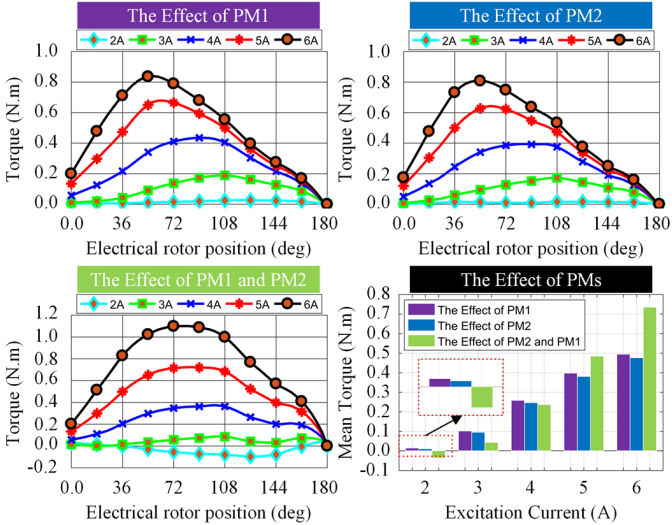

Fig. 11. The contribution of embedded PMs to the torque production.

### E. *Dynamic Simulations*

Fig. 12 displays the voltage, current, and torque waveforms for the four motors at a current hysteresis of 6 A and a nominal speed of 600 rpm. The values of α=1.025$^{om}$ and β=0.326$^{om}$ are extracted from the torque-angle characteristics. The hysteresis band is set to δ=0.2 A, leading to a current ranging from 5.8 A to 6.2 A. The turn-on angle exceeds 180 electrical degrees ($\theta_{on}$=10.7 mechanical degrees), while the turn-off angle is less than 180 electrical degrees ($\theta_{off}$=9.3 mechanical degrees). The torque ripples of Motors 1, 2, 3, and 4 are 127%, 135%, 135%, and 121%, respectively. The dynamic mean torque of Motor 4 is 188.70% higher than that of Motor 1 due to the impact of embedded PMs. As shown in Fig. 12, the torque remains consistently positive, showing the effectiveness of the proposed self-starting and control strategy. Table III summarizes the dynamic performance of the four studied motors, with Motor 4 showing the best dynamic performance. Core and copper losses are nearly equivalent across all motors; however, since Motor 4 yields higher output torque, it achieves the highest efficiency.

### F. *Comparison with the Conventional SRM and other HRMs*

The superior performance of Motor 4 compared to the conventional 8/12 SRM and various HRMs is highlighted in this study. Detailed static results of this comparison are provided in Table IV, demonstrating the clear superiority of Motor 4 over the conventional SRM. Conversely, Motor 1 exhibits inferior performance compared to the conventional SRM, while the other two motors perform nearly similarly. Furthermore, the performance of Motor 4 is benchmarked against the presented

#### TABLE III
#### PREDICTED DYNAMIC BEHAVIOR OF THE MOTORS

| Parameter | Motor 1 | Motor 2 | Motor 3 | Motor 4 |
|---|---|---|---|---|
| Motor volume *(mL)* | 320 | 320 | 320 | 320 |
| Nominal speed, $n_r$ *(rpm)* | 600 | 600 | 600 | 600 |
| RMS phase current *I (A)* | 4.11 | 4.09 | 4.02 | 4.14 |
| Mean Torque $T_{mean}$ *(N.m)* | 0.363 | 0.801 | 0.779 | 1.048 |
| Output power $P_d$ *(W)* | 22.80 | 50.32 | 48.94 | 65.84 |
| Copper losses $P_{Cu}$ *(W)* | 7.37 | 7.28 | 6.96 | 7.51 |
| Total core loss $P_c$ *(W)* | 0.96 | 0.88 | 0.89 | 0.86 |
| Total losses *(W)* | 8.33 | 8.16 | 7.85 | 8.37 |
| Input power $P_{in}$ *(W)* | 31.13 | 58.48 | 56.79 | 74.21 |
| Torque density *(N.m/L)* | 1.134 | 2.503 | 2.434 | 3.275 |
| Torque ripple *(%)* | 127 | 135 | 135 | 121 |
| Torque per ampere *(N.m/A)* | 0.088 | 0.195 | 0.193 | 0.253 |
| Power per ampere *(W/A)* | 5.547 | 12.303 | 12.174 | 15.903 |
| Efficiency $k_e$ *(%)* | 73.24 | 86.04 | 86.17 | 88.72 |

TABLE IV
COMPARISON OF THE DYNAMIC RESULTS WITH OTHER MOTORS [24], [25], [31], [32].

| Parameter | Motor 1 | Motor 2 | Motor 3 | Motor 4 | 8/12 SRM | MT-HESRM [31] | MT-SRM [31] | PM-SRM [32] | HRM [29] | HRM [35] | HRM [36] |
|---|---|---|---|---|---|---|---|---|---|---|---|
| Motor volume with frame (mL) | 320 | 320 | 320 | 320 | 320 | 320 | 320 | 320 | 320 | 320 | 320 |
| PMs volume (mL) | - | 2 | 4 | 6 | - | 2.16 | - | 4.32 | 6.21 | 5.67 | 4.80 |
| RMS Phase current (A) | 4.11 | 4.09 | 4.02 | 4.14 | 4.03 | 2.37 | 3.05 | 3.48 | 2.96 | 2.86 | 2.60 |
| Mean torque (N.m) | 0.363 | 0.801 | 0.779 | 1.048 | 0.882 | 0.67 | 0.41 | 0.980 | 0.480 | 0.540 | 0.520 |
| Output power $P_d$ (W) | 22.80 | 50.32 | 48.94 | 65.84 | 55.41 | 63.14 | 38.64 | 20.52 | 48.10 | 41.50 | 54.40 |
| Torque density (N.m/L) | 1.134 | 2.503 | 2.434 | 3.275 | 2.756 | 2.10 | 1.28 | 3.062 | 1.500 | 1.687 | 1.625 |
| Torque ripple (%) | 127 | 135 | 135 | 121 | 141 | 119 | 121 | 122 | 154 | 133 | 142 |
| Torque per PMs volume (N.m/L) | - | 400.50 | 194.75 | 174.66 | - | 310.18 | - | 226.85 | 77.29 | 95.23 | 108.33 |
| Torque per ampere (N.m/A) | 0.088 | 0.195 | 0.193 | 0.253 | 0.218 | 0.282 | 0.134 | 0.281 | 0.162 | 0.188 | 0.200 |
| Power per ampere (W/A) | 5.547 | 12.303 | 12.174 | 15.903 | 13.749 | 26.641 | 12.668 | 5.896 | 16.250 | 14.510 | 20.92 |
| Efficiency $k_e$ (%) | 73.24 | 86.04 | 86.17 | 88.72 | 84.15 | 80.70 | 60.62 | 36.5 | 83.87 | 76.78 | 85.82 |

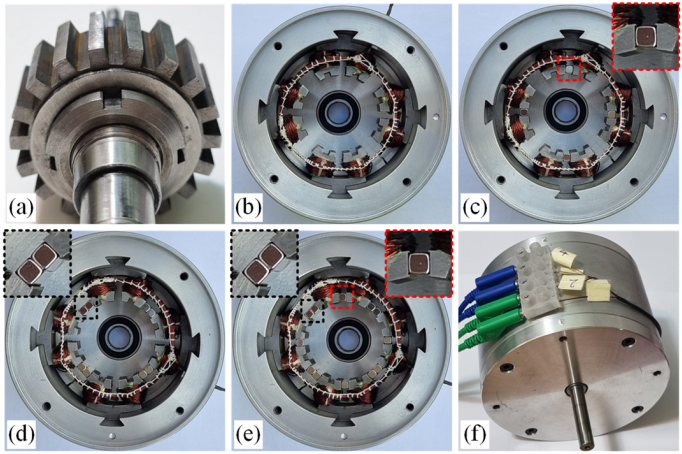

Fig. 13. Prototyped motors: (a) rotor, (b) stator of Motor 1, (c) stator of Motor 2, (d) stator of Motor 3, (e) stator of Motor 4, and (f) assembled Motor.

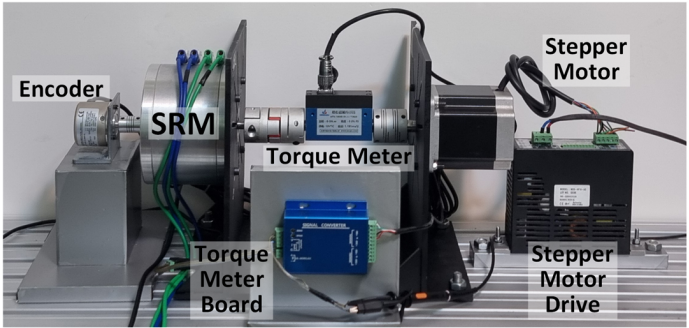

Fig. 14. The setup for measuring static torque angles.

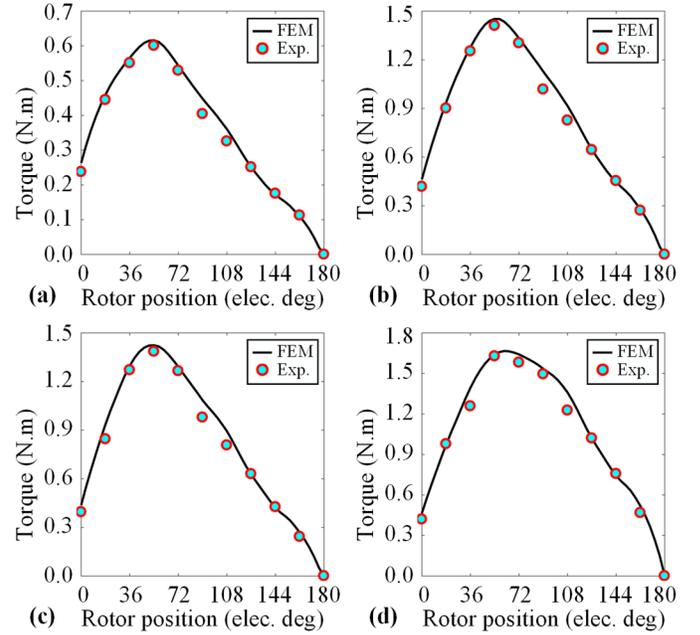

Fig. 15. Measured torque-angle characteristics: (a) Motor 1, (b) Motor 2, (c) Motor 3, and (d) Motor 4.

small appendices secure these modules in place and introduce only a minimal increase in manufacturing costs.

### B. Experimental Static Results

Fig. 14 depicts the setup utilized for measuring the torque-angle characteristics of the motor. It includes a rotary encoder, a torque meter, and a stepper motor to rotate the SRMs at the desired position. These static torque values are recorded at intervals of ten mechanical degrees. In Fig. 15, the torque-angle characteristics of the four motors at a phase current of 6 A are illustrated, showing a great correlation with FEM. The superiority of Motor 4 is also clear.

### C. Experimental Dynamic Results

The dynamic test setup is depicted in Fig. 16. A DC generator, equipped with an adjustable armature resistor and excitation current, serves as the load. The input DC voltage is set at 150 V with a hysteresis current of 6 A. Fig. 17 presents the waveforms of measured voltage, current, and input power at the nominal speed of 600 rpm. The RMS phase current for Motors 1, 2, 3, and 4 measures approximately 4.06 A, 3.98 A, 4.06 A, and 4.09 A, respectively. Table V compares the predicted and measured dynamic performance of the motors, highlighting that Motor 4 achieves higher mean torque, output power, torque density, and efficiency while exhibiting lower torque ripple

HRMs in Table IV, showing its dominance in overall PM performance. Notably, Motor 4 shows significant improvements across several key performance metrics, including mean torque, output power, torque density, and efficiency, compared to both the conventional SRM and other HRMs. Additionally, Motor 3 outperforms the SRMs and all HRMs in terms of torque per PM volume. In conclusion, Motor 4 surpasses the other motors in all indices. Furthermore, comparisons indicate that Motor 4 exhibits the lowest torque ripple.

## IV. PROTOTYPED MOTORS AND EXPERIMENTAL RESULTS

In this section, the motors are prototyped and tested.

### A. Prototyped Motors

Fig. 13 illustrates the prototype of the studied motors. In Motors 2, 3, and 4, PMs are placed within the designated slots between the teeth of the C-cores, enhancing the overall mechanical stability of the stator structure. Additionally, four compared to the other motors. The total losses among the motors are nearly identical, yet due to its superior mean torque and output power, Motor 4 achieves higher efficiency. The calculated efficiencies for Motors 1, 2, 3, and 4 are approximately 73.20%, 86.02%, 86.14%, and 88.66%, respectively.

TABLE V
COMPARISON OF PREDICTED DYNAMIC BEHAVIOR AND MEASURED RESULTS FOUR TWO-PHASE PROPOSED MOTORS.

| Parameter | Motor 1 | | | Motor 2 | | | Motor 3 | | | Motor 4 | | |
|---|---|---|---|---|---|---|---|---|---|---|---|---|
| | Predicted | Measured | \|Error (%)\| | Predicted | Measured | \|Error (%)\| | Predicted | Measured | \|Error (%)\| | Predicted | Measured | \|Error (%)\| |
| RMS phase current $I$ (A) | 4.11 | 4.06 | 1.21 | 4.09 | 3.98 | 2.69 | 4.02 | 4.06 | 0.99 | 4.14 | 4.09 | 1.20 |
| Mean Torque $T_{mean}$ (N.m) | 0.363 | 0.347 | 4.40 | 0.801 | 0.767 | 4.24 | 0.779 | 0.749 | 3.85 | 1.048 | 1.000 | 4.58 |
| Output power $P_d$ (W) | 22.80 | 21.80 | 4.38 | 50.32 | 48.19 | 4.23 | 48.94 | 47.06 | 3.84 | 65.84 | 62.83 | 4.57 |
| Total losses (W) | 8.33 | 7.98 | 4.20 | 8.16 | 7.83 | 4.04 | 7.85 | 7.57 | 3.56 | 8.37 | 8.03 | 4.06 |
| Input power $P_{in}$ (W) | 31.13 | 29.78 | 4.33 | 58.48 | 56.02 | 4.20 | 56.79 | 54.63 | 3.80 | 74.21 | 70.86 | 4.51 |
| Torque per ampere (N.m/A) | 0.088 | 0.085 | 3.41 | 0.195 | 0.192 | 1.53 | 0.193 | 0.184 | 4.66 | 0.253 | 0.244 | 3.55 |
| Power per ampere (W/A) | 5.547 | 5.369 | 3.21 | 12.303 | 12.108 | 1.58 | 12.174 | 11.591 | 4.79 | 15.903 | 15.362 | 3.40 |
| Efficiency $k_e$ (%) | 73.24 | 73.20 | 0.05 | 86.04 | 86.02 | 0.02 | 86.17 | 86.14 | 0.03 | 88.72 | 88.66 | 0.06 |

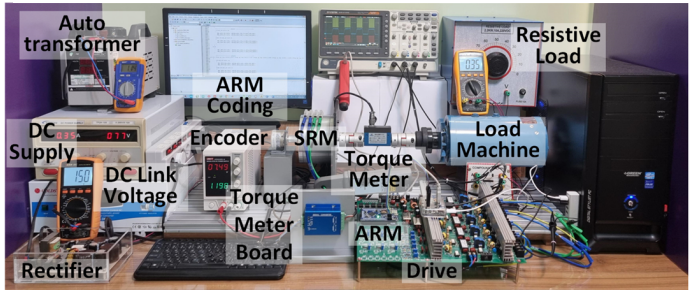

Fig. 16. Drive and control setup.

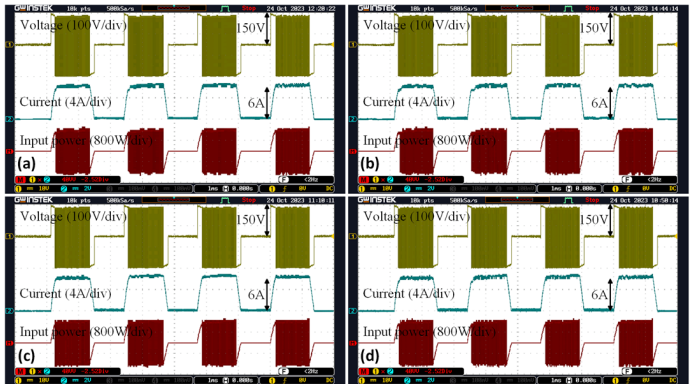

Fig. 17. Measured voltage, current, and input power of the motors at 600 rpm: (a) Motor 1, (b) Motor 2, (c) Motor 3, and (d) Motor 4.

## V. CONCLUSION

A double-teeth C-core SRM with embedded PMs, coupled with a novel drive technique to enhance self-starting capabilities, is examined. Magnetic equivalent circuits are developed for the design and analysis of the investigated SRMs. Three PM placement topologies, alongside a conventional SRM, are explored to determine the most effective PM placement strategy. FEM is utilized in the analysis and design processes. The four SRMs are prototyped and tested, and the experimental results show a strong correlation with FEM simulations. The effectiveness of the PM-assisted SRMs in enhancing torque density and efficiency is demonstrated. It is shown that PMs in set 1 are more effective than set 2, as a greater portion of the flux generated by PMs of set 1 enters the air-gap, compared to that of PMs in set 2. The performance of Motor 4 shows superiority over the conventional SRM and various HRMs. The proposed drive strategy effectively eliminates the negative torque generated by the self-starting technique implemented in the topology of the SRMs.

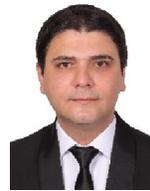

**Gholamreza Davarpanah** (M'21) received his B.Sc. degree with distinction (ranking first) from the Islamic Azad University, Bojnord Branch, Iran, in 2009, specializing in electrical engineering (electronics). Subsequently, he completed his M.Sc. degree at Amirkabir University of Technology, Tehran, in 2014, with a focus on electrical engineering (electrical machines and power electronics). Currently serving as a Research Assistant at the Center of Excellence on Applied Electromagnetic Systems, School of Electrical and Computer Engineering, College of Engineering, University of Tehran, Tehran, his primary research areas encompass electric machine design, optimization, modeling, and prototyping. His expertise extends to switched reluctance machines, permanent magnet machines, VR motors, power electronics, and drives.

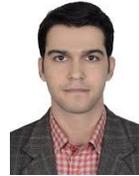

**Sajjad Mohammadi** (S'13) received a B.S. degree (honors) from the Kermanshah University of Technology, Kermanshah, Iran, in 2011 and an M.S. degree (honors) from the Amirkabir University of Technology, Tehran, in 2014, all in electrical engineering. He also got an M.Sc. degree and a Ph.D. degree in electrical engineering and computer science from the Massachusetts Institute of Technology (MIT) in 2018 and 2021, respectively. His research interests include energy conversion, design and control of electromechanical devices and electric machines, drives, power electronics, power systems, and energy policy.

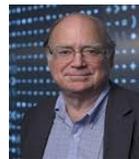

**James L. Kirtley, Jr.** (F'90) received a Ph.D. degree from the Massachusetts Institute of Technology (MIT), Cambridge, MA, USA, in 1971. He is currently a Professor of Electrical Engineering at MIT. He has been with the Large Steam Turbine Generator Department, General Electric, Schenectady, NY, USA, as an Electrical Engineer, with Satcon, Boston, MA, USA, Technology Corporation as the Vice President and General Manager of the Tech Center and as the Chief Scientist, and with the Swiss Federal Institute of Technology, Zürich, Switzerland, as a Gastdozent. He continues as a Director of Satcon. He is a Specialist in electric machinery and electric power systems. Dr. Kirtley received the IEEE Third Millenium Medal in 2000 and the Nikola Tesla Prize in 2002. He served as the Editor-in-Chief of the IEEE Transactions on Energy Conversion from 1998 to 2006 and continues to serve as an Editor for that journal and as a member of the Editorial Board of the Electric Power Components and Systems journal. He was elected to the United States National Academy of Engineering in 2007. He is a Registered Professional Engineer in Massachusetts.